\documentclass[aps,prl,twocolumn,amsmath,floatfix,superscriptaddress,showpacs]{revtex4}
\usepackage{graphicx}
\usepackage{bm}
\usepackage{epstopdf}
\usepackage{dcolumn}
\usepackage{undertilde}
\usepackage{color}

\begin{document}

\title{The Affine-Nonaffine Transition in Anisotropic Semiflexible Networks}
\author{Andrew R. Missel}
\email[]{missel@ucla.edu}
\affiliation{Department of Chemistry and Biochemistry, UCLA, Los Angeles, CA 90095}
\author{Mo Bai}
\affiliation{Department of Mechanical and Aerospace Engineering, UCLA, Los Angeles, CA 90095}
\author{William S. Klug}
\affiliation{Department of Mechanical and Aerospace Engineering, UCLA, Los Angeles, CA 90095}
\affiliation{California Nanosystems Institute, UCLA, Los Angeles, CA 90095}
\author{Alex J. Levine}
\affiliation{Department of Chemistry and Biochemistry, UCLA, Los Angeles, CA 90095}
\affiliation{California Nanosystems Institute, UCLA, Los Angeles, CA 90095}

\date{\today}

\begin{abstract}
We study the mechanics of nematically ordered  semiflexible networks showing that they, like isotropic networks, undergo an affine to non-affine cross-over controlled by the ratio of the filament length to the non-affinity length. Deep in the non-affine regime, however, these anisotropic networks exhibit a much more complex mechanical response characterized by a vanishing linear response regime for highly ordered networks and a generically more complex dependence of the shear modulus upon the direction of shear relative to the nematic director.  We show that these features can be understood in terms of a generalized floppy modes analysis of the non-affine mechanics and a type of cooperative Euler buckling. 
\end{abstract}


\maketitle

The mechanical properties of living eukaryotic cells are controlled by a low-density cross-linked network of semiflexible protein filaments~\cite{alberts2002molecular}.  These filaments, composed  primarily of F-actin, are densely cross-linked on the scale of their thermal persistence length. Understanding the material properties of such {\em semiflexible} gels is not only a forefront problem in biophysics but also presents a broader challenge in polymer physics.  These cytoskeletal semiflexible networks can support elastic stress in both filament stretching and bending. This is in contradistinction to typical polymeric gels for which rubber elasticity theory applies~\cite{rubinstein2003polymer}. Those {\em flexible} polymer systems, where the thermal persistence length of the constituent filaments is significantly smaller than the mean distance between cross-links along a given filament, store elastic energy only in the stretching of filament random walks between consecutive cross-links. 

Previous work~\cite{head2003deformation,heussinger2006floppy} has shown that, as a function of increasing network density or filament bending stiffness, semiflexible networks admit a sharp cross-over from a bending dominated elastically compliant regime that sets in for networks at densities above the rigidity percolation transition~\cite{thorpe1983continuous}, to stiffer networks where elastic strain energy is stored primarily in the stretching of filaments. Moreover, under uniformly applied strain at the boundaries, the geometry of the deformation field in the softer, bending dominated regime is  spatially heterogeneous (i.e., non-affine) and shows large deviations from the affine deformation prediction of continuum elasticity theory~\cite{landau1995theory}.  This cross-over is controlled by the ratio of the filament length $L$ to the so-called non-affinity length $\lambda$. 

This earlier work concentrated on the mechanics of statistically isotropic filament networks. Cytoskeletal networks however, are dominated by oriented stress fibers and typically have lower symmetry. Motivated by this point we explore in this Letter the mechanics of anisotropic networks by considering random networks with a nonuniform probability distribution of filament orientations. Although these anisotropic networks, characterized by both a filament density and nematic order parameter, undergo a non-affine to affine cross-over that is also controlled by $L/\lambda$, we find that the mechanics of anisotropic semiflexible networks are considerably more complex. Specifically, we find that: (i) their linear response regime narrows dramatically at high nematic order,  and (ii) their shear response in the non-affine regime develops a complex dependence upon the angle between the shearing direction and the nematic direction that cannot be understood in terms of continuum linear elasticity.  We can account for both of these results using a self-consistent model of cooperative Euler buckling closely related to the ``floppy modes'' analysis of non-affine mechanics by Heussinger and Frey~\cite{heussinger2006floppy}.  The analysis of anisotropic semiflexible networks more generally provides deeper insight into the mechanics of the non-affine regime.  In spite of the greater apparent complexity of anisotropic networks, their order makes them more amenable to analysis than the isotropic networks where the affine/non-affine cross-over was discovered.

\begin{figure}[htb]
\includegraphics[width=\columnwidth]{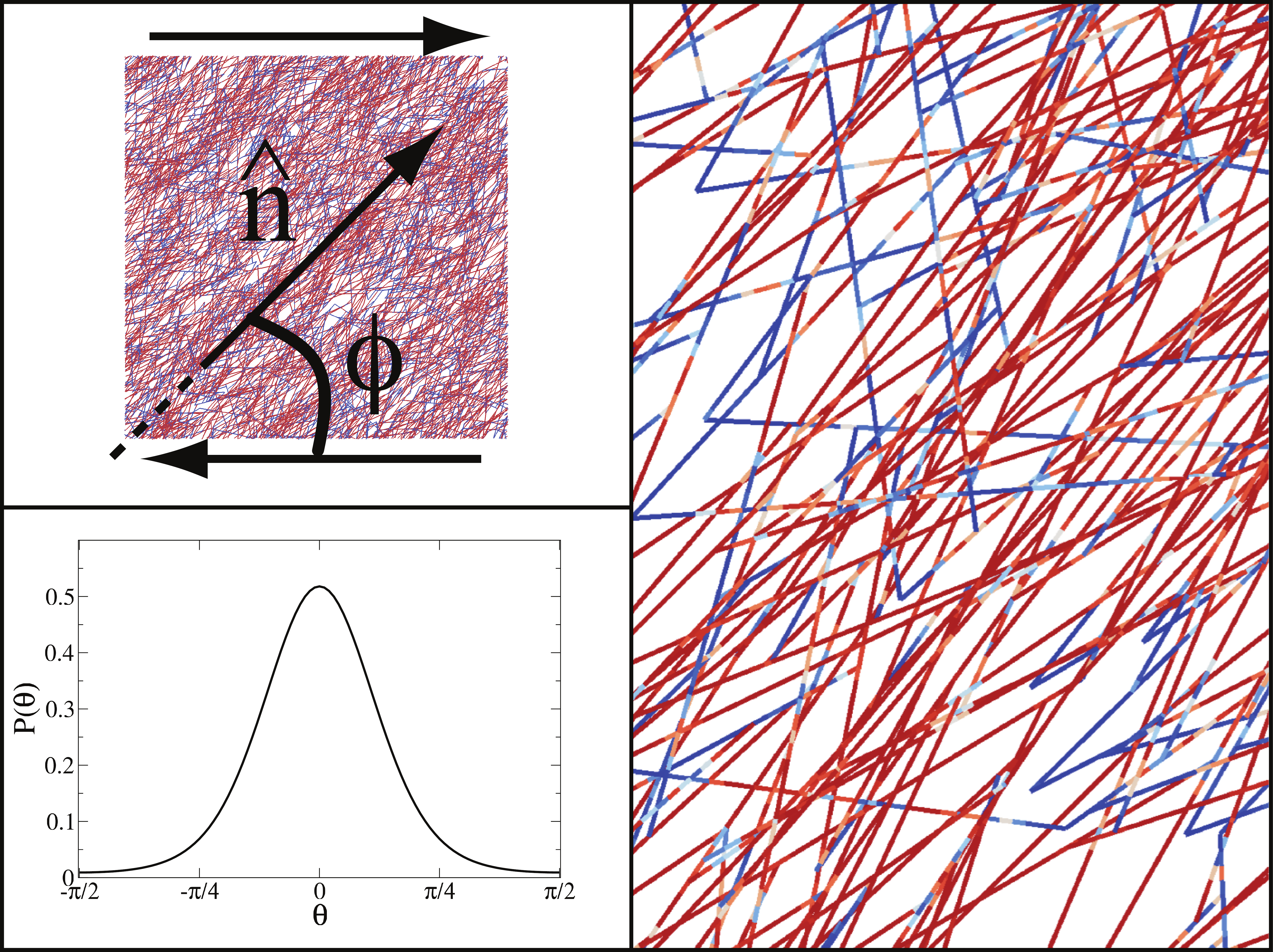}
\caption{\label{network}Top left: A typical nematic network with diagram defining the shear orientation angle $\phi$.  Bottom left: The angular distribution $P(\theta)$ of filament orientations relative to the nematic director.  Right: A closeup of a network (shear direction indicated by arrows) showing the partitioning of elastic energy into filament stretching (red) and bending (blue).}
\end{figure}
We began by numerically studying two-dimensional filament networks composed of identical filaments of length $L$ and having bending and stretching moduli $\kappa$ and $\mu$ respectively.  To produce a system of $N$ filaments with statistically uniform density and filament anisotropy characterized by nematic order parameter $S$, one endpoint of each filament was laid down at random in a square box of edge length $W$ and area $W^{2} = A$, and assigned an orientation chosen from the distribution $P(\theta)$ with respect to nematic direction $\hat{n}$.  Crossing points between filaments were treated as incompliant crosslinks. Filaments were added until the system reached the selected density.  A representative example of such a network, a magnified region thereof, and the distribution of filament angles are shown in Fig.~(\ref{network}).  The filaments were discretized such that nodes were placed at all crosslinks and at regular intervals between crosslinks to allow for bending.  The energy of the network is given by
\begin{equation}
\mathcal{H} = \frac{\mu}{2}\sum_{\text{segments}}\frac{\left(s-s_0\right)^2}{s_0} + \frac{\kappa}{2}\sum_{\text{angles}}\frac{1-\cos\beta}{\ell_0},
\end{equation}
where $s$ is the length of a segment of rest length $s_0$, $\beta$ is the angle between adjacent segments on one filament, and $\ell_0$ is the average rest length of the two segments surrounding an angle spring.  We used the Lees-Edwards method~\cite{lees1972computer} to shear the system with periodic boundary conditions. 

The networks are characterized by a nematic order parameter $S$ and a density, measured in terms of the mean distance between cross-links along a filament, $\ell_{c}$.  The mechanics of the monodisperse filaments is set by $\ell_{b} = \sqrt{\kappa/\mu}$.  Finally, all energy scales are set by a single filament elastic modulus; we measure all moduli in terms of $\mu$. 

For a given $P(\theta)$, the nematic order parameter $S=\int_{-\pi}^{\pi}d\theta\,P(\theta)\cos2\theta$ describes the degree of anisotropy; $S=0$ corresponds to an isotropic system, and $S=1$ is a system with all filaments aligned.  We compute the mean distance between cross-links from the filament density $\rho = N/A$ as follows: Consider a  straight filament lying at an angle $\theta$ with respect to the nematic director. Given a second filament at some angle $\psi$ with respect to the first, the probability of their crossing is $L^2|\sin\psi|/A$. Integrating over $\psi$ with a weight $P(\theta+\psi)$ we find that the mean number of filaments crossing the original one lying at angle $\theta$ with respect to $\hat{n}$ is Poisson distributed with mean
\begin{equation}
n_c(\theta) = 2\rho L^2\int_0^{\pi}d\psi\,\sin\psi\,P(\theta+\psi).
\end{equation}
Integrating over the orientation of that original filament, we find that the mean distance between cross-links is given by
\begin{equation}
\ell_c = L\,\frac{1+\left\langle e^{-n_c}\right\rangle-2\left(\left\langle\frac{1}{n_c}\right\rangle-\left\langle\frac{e^{-n_c}}{n_c}\right\rangle\right)}{\left\langle n_c\right\rangle-\left(1-\left\langle e^{-n_c}\right\rangle\right)},
\end{equation}
where the angle brackets denote averaging over $\theta$ with weight $P(\theta)$.  For an isotropic gel, Head \emph{et. al} found that non-affine to affine cross-over is controlled by $L/\lambda$ where $\lambda = \ell_{c}(\ell_{c}/\ell_{b})^{z}$ with $z=1/3$~\cite{head2003deformation}. Scaling arguments~\cite{head2003deformation} and mean field theories~\cite{das2007effective} suggest $z=2/5,1/4$ respectively.   

Two-dimensional elastic continua with nematic order may be characterized by four independent elastic constants \cite{mohazzabi1998elastic}.  Writing the elastic constant tensor in the usual way, $E = \frac{1}{2}\int d^{2}\bm{x}\,C_{ijkl}u_{ij}u_{kl}$~\cite{chaikin2000principles}, we can identify these as $C_{1111}$, $C_{2222}$, $C_{1212}$, and $C_{1122}$. Assuming affine deformation, these quantities are all linearly proportional to the stretching modulus of the constituent filaments via the relation 
\begin{equation}
\label{elastic-constants}
C_{ijkl} = \mu\rho L\int_{0}^{2\pi}d\theta\,P(\theta)\hat{e}_i\hat{e}_j\hat{e}_k\hat{e}_l\sum_{n=2}^{\infty}p_n(\theta)L\frac{n-1}{n+1},
\end{equation}
where $\hat{e}(\theta)$ is the direction along a filament oriented at angle $\theta$ and $n$, the number of cross-links per filament, is summed over using their Poisson distribution $p_n(\theta)$. We focus exclusively on the shear modulus $G(\phi)$, which, for a given {\em shear orientation angle} $\phi$, the angle between the displacement and nematic directions as shown in Fig.~\ref{network}, takes the form, using Eq.~(\ref{elastic-constants}), of 
\begin{equation}
G_{\text{affine}}(\phi) = G_{\text{affine}}(0)\left[1-8\sin^2\phi\cos^2\phi\right]+\Gamma \sin^2\phi\cos^2\phi\,
\end{equation}
with
\begin{equation}
\Gamma = \mu\rho L\int_0^{2\pi}d\theta\,P(\theta)\sum_{n=2}^{\infty}p_n(\theta)\frac{n-1}{n+1}.
\end{equation}
We refer to the above as the affine prediction for the shear modulus of our anisotropic solids.  The angular dependence of $G_{\text{affine}}(\phi)$ has a simple interpretation. A shear deformation with displacement along $\hat{x}$ is, to linear order, equivalent to stretching and compressing 
along the $\pm \pi/4$ directions with respect to $\hat{x}$.  Filaments oriented at angles $\pm\pi/4$ with respect to $\hat{x}$  will be 
stretched/compressed along their axes, and thus have the most strain energy.  Thus networks sheared so that the displacement and nematic director are separated by these angles will have maximal shear moduli.   

The depression of the shear modulus $G$  below its affine value $G_{\text{affine}}$ serves as a mechanical measure of the departure from affinity. 
Geometric measures of this departure~\cite{head2003deformation,didonna2005nonaffine} measure the spatial heterogeneity of the strain 
field  as  a function of length scale. Following Head et al.~\cite{head2003deformation} we adopt  as our geometric measure of nonaffinity 
$\Delta(r) = \langle\left(\omega-\omega_{\text{affine}}\right)^2\rangle(r)$, where $\omega$ is the strain-induced change in angle between 
an arbitrary axis and the line joining two network points separated by a distance $r$, $\omega_{\text{affine}}$ is the change of this angle given purely affine deformations, and the average is taken over all point pairs with separation $r$ in a network.
\begin{figure}[htb]
\includegraphics[width=\columnwidth]{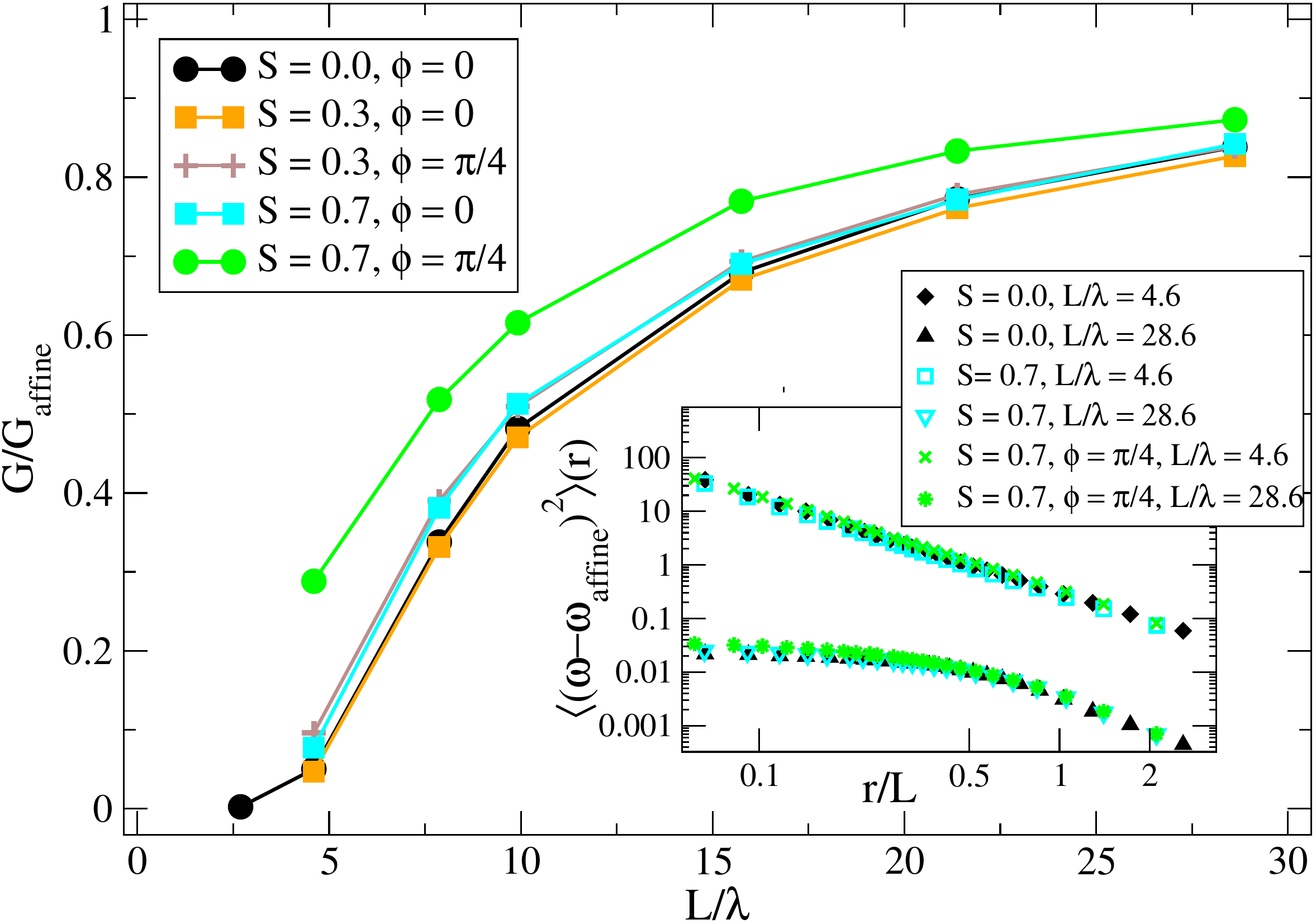}
\caption{\label{ggaff}Normalized shear modulus $G/G_{\text{affine}}$ versus $L/\lambda$ for varying degrees of anisotropy $S$ and shear orientation angle $\phi$. For highly anisotropic gels non-affine networks is there a large shear orientation angle dependence to the deviation from the affine shear modulus prediction.  Inset: The geometric nonaffinity measure $\Delta$ as a function of length scale for six different cases, including the $S=0.7$ network at various values of $\phi$, showing that $\Delta$, unlike the modulus, is insensitive to the shearing angle. Moduli measured at $\gamma = 10^{-3}$.}
\end{figure}

Fig.~\ref{ggaff} plots the normalized shear modulus $G/G_{\text{affine}}$ as a function of $L/\lambda$ for various values of $S$ and $\phi$.  These data show reasonable collapse onto a universal curve, demonstrating that the affine to non-affine cross-over persists for anisotropic networks. The shear modulus of presumably affine networks ($L/\lambda> 15$) nearly saturates the affine prediction and has a shear orientation angle dependence as predicted by the affine theory. This can be seen seen by the collapse of the ratios $G/G_{\text{affine}}$ for various values  of $\phi$ in the affine regime. Deep in the non-affine regime, however, the measured shear modulus of highly anisotropic networks ($S=0.7$) varies more strongly with $\phi$ than can be understood by the affine theory. Moreover, this additional angle dependence breaks the universal data collapse observed in isotropic and more weakly anisotropic networks. This breakdown of continuum linear elasticity is also seen in the divergence of $\Delta(r)$ 
as $r \to 0$ (inset)~\cite{head2003deformation}. This geometric measure is {\em independent of the shearing angle.}  Thus, the degree of heterogeneity of the deformation field depends solely on $L/\lambda$, but the mechanics of the anisotropic networks in the non-affine regime appears to have a novel and more complex dependence on $\phi$.   

\begin{figure}[htb]
\includegraphics[width=\columnwidth]{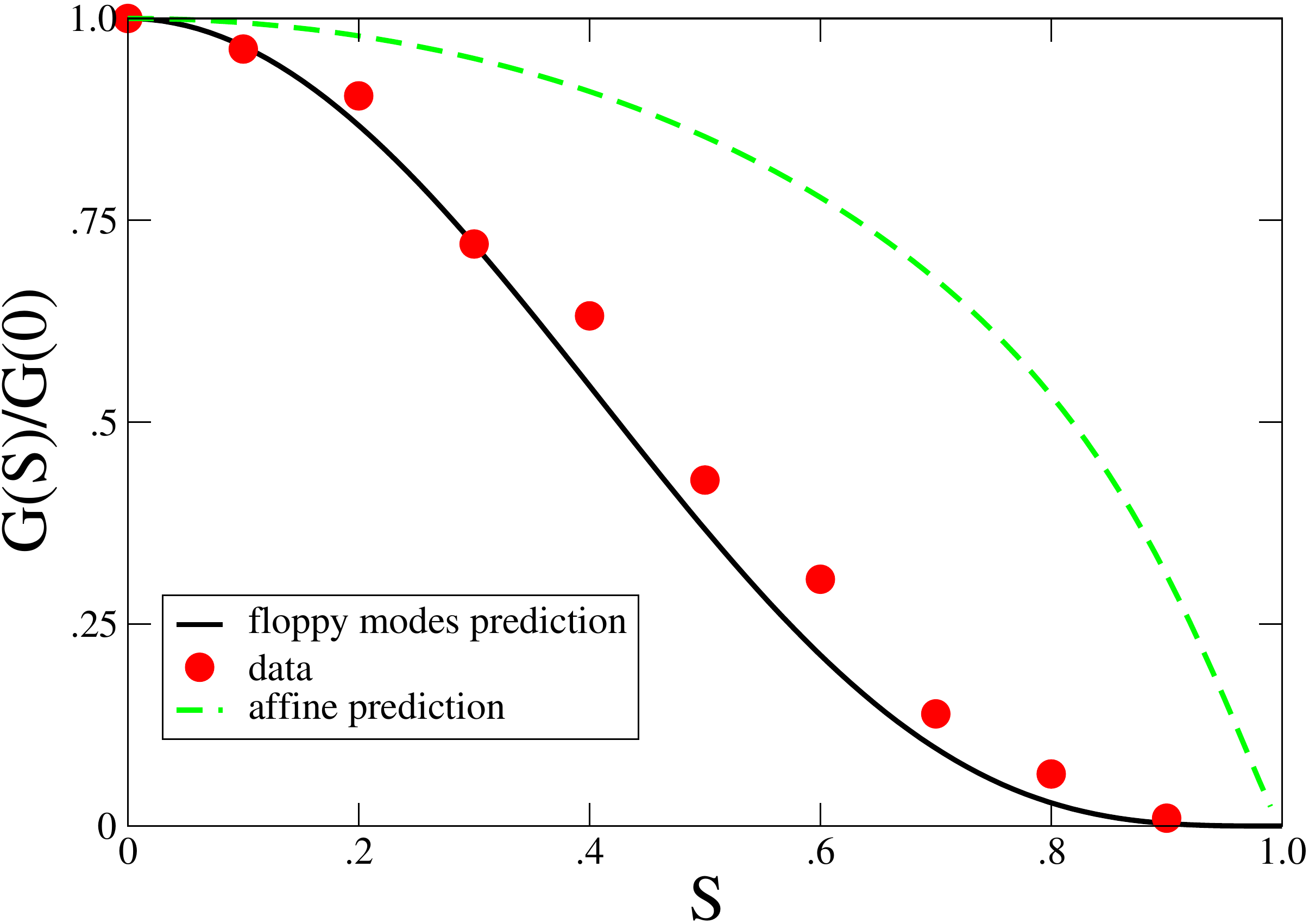}
\caption{\label{constdens} Shear modulus as a function of the nematic order parameter $S$ (for fixed filament density) normalized by its value at $S=0$ 
and $L/\lambda \simeq 5$ .}
\end{figure}
To understand the mechanics of anisotropic networks deep in the non-affine regime, we turn to an extension of the ``floppy modes'' model of Heussinger \emph{et.\ al}~\cite{heussinger2006floppy}.  This model assumes that the deformation energy is stored predominantly in filament bending. For a filament segment of  length $\ell$ this energy is  $E_{\text{bend}}\simeq\kappa\delta_{\text{na}}^2/\ell^3$, where $\delta_{\text{na}}$ is the typical size of axial displacements.  For larger values of $\ell$, it is energetically favorable for segments to bend, thus inducing a small amount of bending on connecting filaments; however, below some length $\ell_{\text{min}}$, the energetic cost of bending becomes too high. The displacement of those filaments results in stresses in their neighbors causing them to bend. One self-consistently determines $\ell_{\text{min}}$ by balancing the bending energy of these small segments with the total bending energy of the sea of longer filaments to which they are connected. This balance determines the elastic energy storage and thus the modulus.

We consider an idealized anisotropic gel with two classes of filaments: (i) {\em nematic}  filaments oriented at $\theta=0$ and (ii) {\em impurity} filaments oriented at $\theta=\pi/2$.  The fraction of nematic filaments $2S-1$ is set to reproduce the correct nematic order parameter.  Now the balance of energy of nematic filaments with their impurity neighbors and impurity energy with nematic neighbors results in two equations that must be solved simultaneously:
\begin{eqnarray}
\frac{\kappa\delta_{\text{na}}^2}{\ell_{\text{min,N}}^3} & = & n_{c,I}\int_{\ell_{\text{min,I}}}^{\infty}d\ell_{I}\,P(\ell_I)\frac{\kappa\delta_{\text{na}}^2}{\ell_I^3}\qquad\text{and}\nonumber\\
\frac{\kappa\delta_{\text{na}}^2}{\ell_{\text{min,I}}^3} & = & n_{c,N}\int_{\ell_{\text{min,N}}}^{\infty}d\ell_{N}\,P(\ell_N)\frac{\kappa\delta_{\text{na}}^2}{\ell_N^3},
\end{eqnarray}
where $P(\ell)$ is the distribution of segment lengths nematic (N)  and impurity (I)  filaments.  The solution, as in the isotropic case, gives $G\sim\rho^7$, but also predicts the dependence of the shear modulus on $S$ deep in the non-affine phase.  In Fig.~\ref{constdens} the anisotropic floppy mode theory (sold line) agrees well with the simulation data (red dots), while the affine prediction (green dashed line) does not.  
\begin{figure}[htb]
\includegraphics[width=\columnwidth]{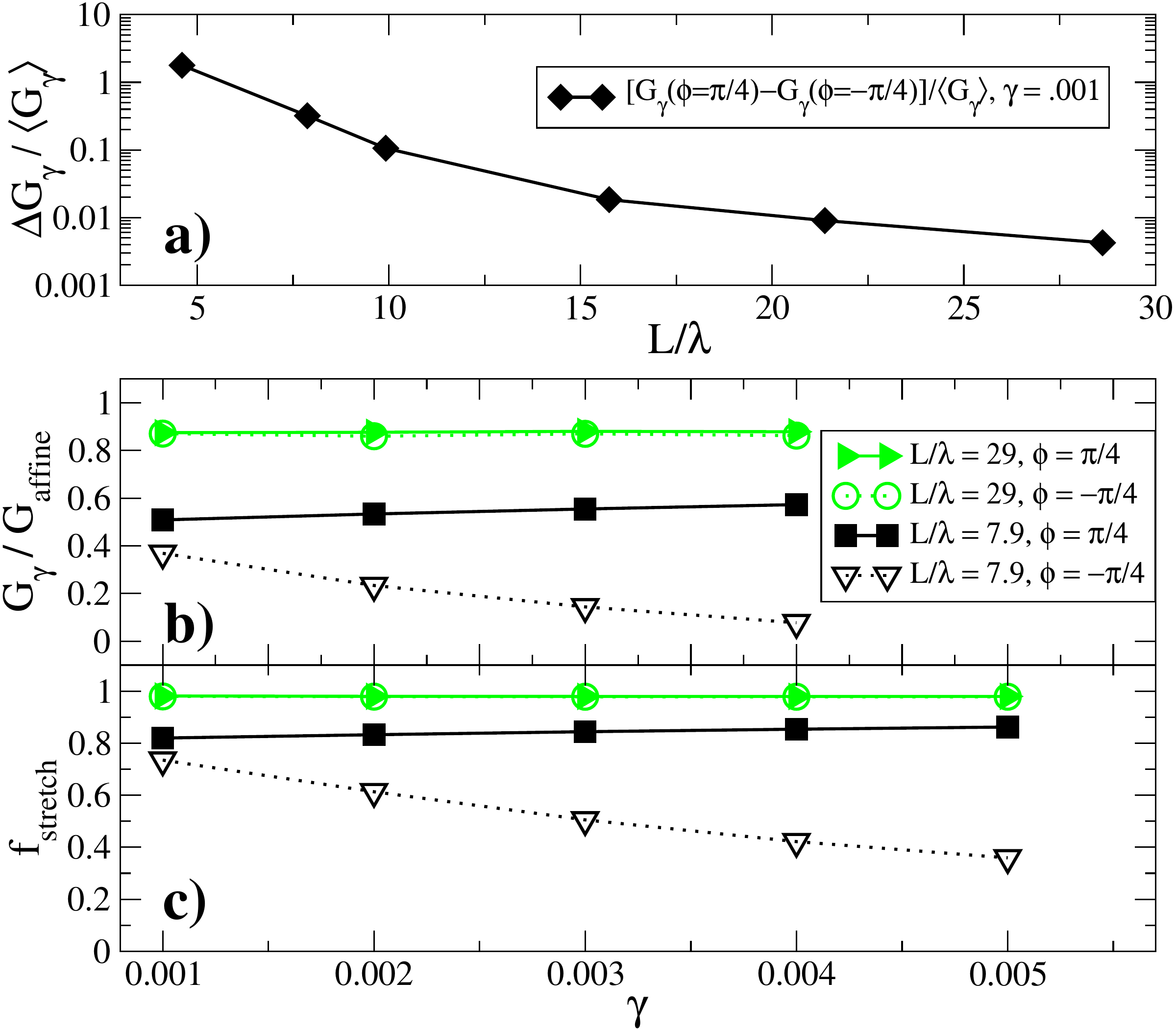}
\caption{\label{soften}(a) Difference in $G$ for $\phi=\pm\pi/4$ at $S=.7$ as a function of $L/\lambda$.  The shear modulus was measured strain $\gamma=.001$; (b) $G/G_{\text{affine}}$ vs. $\gamma$ for $S=.7$ for various values of $L/\lambda$ and $\phi$; (c) Fraction of energy in stretching as a function of shear for the same networks.}
\end{figure}

This result does not explain the unexpectedly stronger $\phi$ dependence in the non-affine regime. To account for this effect, we must turn to the \emph{nonlinear} response of anisotropic networks in the non-affine regime.  Fig.~\ref{soften}(a) plots the difference between $G$ measured at $\phi=\pi/4$ and $\phi=-\pi/4$ at $\gamma = 0.001$ as a function of $L/\lambda$ for a network with $S=0.7$.  $G_{\gamma}$ is computed as the second derivative of the energy density with respect to $\gamma$.   As these two orientations are equivalent under a mirror reflection, according to continuum linear elasticity $\Delta G_{\gamma}$ should vanish.  Indeed it does so as we reach the affine regime: $L/\lambda \gg 1$. For non-affine networks, however, $\Delta G_{\gamma}$ becomes large.  Examining Fig.~\ref{soften}(b) we see the origin of this particular deviation from continuum linear elasticity for non-affine gels ($L/\lambda = 7.9$).  $\Delta G_{\gamma}$ increases with $\gamma$, demonstrating that it is an inherently nonlinear effect. Specifically, the modulus for  $\phi = - \pi/4$ (dotted black line) monotonically softens relative to that of $\phi =  \pi/4$  (solid black line)  as $\gamma$ increases.  For an affine network ($L/\lambda = 29$) these two shear moduli (green solid and dotted lines) are equal for all strains explored showing that affine networks have a significantly larger linear response regime.  

The nonlinearity that leads to the difference $\Delta G_{\gamma}$ can be traced to cooperative Euler buckling of the network that leads to the nonlinear softening of the shear modulus at $\phi=-\pi/4$ since, at this network orientation relative to the shearing direction, more filaments are placed under compressive stress. Moreover, these filaments have larger values of $\ell_{c}$ allowing for buckling at smaller strains. To demonstrate this, we plot in Fig.~\ref{soften}(c) the fraction of elastic energy stored in stretching as a function of both strain and $L/\lambda$. In the non-affine regime $L/\lambda=7.9$, there is generically a smaller fraction of elastic energy stored in stretching, as expected from previous work, but the fraction of stretching energy decreases dramatically with strain as the network is sheared at angle $\phi = - \pi/4$ (dotted black line) applying compressive stresses to the filaments along $\hat{n}$.  Shearing along $\phi = \pi/4$ puts the nematically aligned filaments under tension. The filaments now under compression are generically cross-linked on a much finer scale so that the network now collectively resists buckling.  

To estimate the shear at which we expect $\Delta G_{\gamma}>0$, we note that Euler buckling of a filament of length $\ell$ occurs above a critical compressive load $F_{c}= \pi^2\kappa/\ell^2$~\cite{landau1995theory}. For an affine deformation, $F=\mu \gamma/2$, with the 
compressive strain along $\phi=-\pi/4$.  
We find a critical strain $\gamma_c=2\pi^2(\ell_b/\ell)^2$ to cause a  simply-supported segment of length $\ell$ to buckle. 
To find upper and lower bounds for $\gamma_{c}$, we imagine that buckling sets in on scales $\ell_{c} \le \ell \le L$. For $L/\lambda=7.9$ 
(Fig.~\ref{soften}) we find that $3\times10^{-5} \le \gamma_c \le 10^{-2}$ in reasonable agreement with the data.

We have found that anisotropic networks, much like isotropic networks, undergo an affine to non-affine cross-over that is controlled by $L/\lambda$.  The geometric measure of non-affinity depends on 
$L/\lambda$ in a manner independent of anisotropy.  The mechanics of the non-affine regime of anisotropic networks, however, is quite complex, 
characterized by a vanishing linear response regime \cite{wyart2008} in the limit of increasing nematic order parameter. The overall 
dependence of the network's shear modulus upon $S$ can be understood in terms of a generalization of the floppy 
modes picture. The vanishing of the linear response regime reflects the appearance of a type of cooperative Euler 
buckling.  We note that, while affine networks of arbitrary anisotropy can be understood in terms of elastic continua, non-affine and anisotropic networks allow for a much more complex set of mechanics. This result suggests that the mechanics of ordered cytoskeletal structures must be modeled with particular care in the non-affine regime.  Finally, we propose that insights into the highly nonlinear mechanics of anisotropic non-affine networks should lead toward a more complete theory of the mechanics of the non-affine regime, particularly in the nonlinear regime. 

\begin{acknowledgements}
The authors gratefully acknowledge support for this work from NSF-CMMI-0800533.
\end{acknowledgements}

\begin{thebibliography}{12}
\expandafter\ifx\csname natexlab\endcsname\relax\def\natexlab#1{#1}\fi
\expandafter\ifx\csname bibnamefont\endcsname\relax
  \def\bibnamefont#1{#1}\fi
\expandafter\ifx\csname bibfnamefont\endcsname\relax
  \def\bibfnamefont#1{#1}\fi
\expandafter\ifx\csname citenamefont\endcsname\relax
  \def\citenamefont#1{#1}\fi
\expandafter\ifx\csname url\endcsname\relax
  \def\url#1{\texttt{#1}}\fi
\expandafter\ifx\csname urlprefix\endcsname\relax\def\urlprefix{URL }\fi
\providecommand{\bibinfo}[2]{#2}
\providecommand{\eprint}[2][]{\url{#2}}

\bibitem[{\citenamefont{Alberts et~al.}(2002)\citenamefont{Alberts, Bray,
  Lewis, Raff, Roberts, and Watson}}]{alberts2002molecular}
\bibinfo{author}{\bibfnamefont{B.}~\bibnamefont{Alberts}},
  \bibinfo{author}{\bibfnamefont{D.}~\bibnamefont{Bray}},
  \bibinfo{author}{\bibfnamefont{J.}~\bibnamefont{Lewis}},
  \bibinfo{author}{\bibfnamefont{M.}~\bibnamefont{Raff}},
  \bibinfo{author}{\bibfnamefont{K.}~\bibnamefont{Roberts}}, \bibnamefont{and}
  \bibinfo{author}{\bibfnamefont{J.}~\bibnamefont{Watson}},
  \emph{\bibinfo{title}{{Molecular Biology of the Cell}}}
  (\bibinfo{publisher}{Garland
  Science, New York},\bibinfo{year}{2002}).

\bibitem[{\citenamefont{Rubinstein and Colby}(2003)}]{rubinstein2003polymer}
\bibinfo{author}{\bibfnamefont{M.}~\bibnamefont{Rubinstein}} \bibnamefont{and}
  \bibinfo{author}{\bibfnamefont{R.}~\bibnamefont{Colby}},
  \emph{\bibinfo{title}{{Polymer Physics}}} (\bibinfo{publisher}{Oxford
  University Press, USA}, \bibinfo{year}{2003}).

\bibitem[{\citenamefont{Head et~al.}(2003{\natexlab{a}})\citenamefont{Head,
  Levine, and MacKintosh}}]{head2003deformation}
\bibinfo{author}{\bibfnamefont{D.}~\bibnamefont{Head}},
  \bibinfo{author}{\bibfnamefont{A.}~\bibnamefont{Levine}}, \bibnamefont{and}
  \bibinfo{author}{\bibfnamefont{F.}~\bibnamefont{MacKintosh}},
  \bibinfo{journal}{Physical Review Letters} \textbf{\bibinfo{volume}{91}},
  \bibinfo{pages}{108102} (\bibinfo{year}{2003}{\natexlab{a}}), \bibinfo{author}{\bibfnamefont{D.}~\bibnamefont{Head}},
  \bibinfo{author}{\bibfnamefont{A.}~\bibnamefont{Levine}}, \bibnamefont{and}
  \bibinfo{author}{\bibfnamefont{F.}~\bibnamefont{MacKintosh}},
  \bibinfo{journal}{Physical Review E} \textbf{\bibinfo{volume}{68}},
  \bibinfo{pages}{61907} (\bibinfo{year}{2003}{\natexlab{b}}), \bibinfo{author}{\bibfnamefont{D.}~\bibnamefont{Head}},
  \bibinfo{author}{\bibfnamefont{A.}~\bibnamefont{Levine}}, \bibnamefont{and}
  \bibinfo{author}{\bibfnamefont{F.}~\bibnamefont{MacKintosh}},
  \bibinfo{journal}{Physical Review E} \textbf{\bibinfo{volume}{72}},
  \bibinfo{pages}{61914} (\bibinfo{year}{2005}).



\bibitem[{\citenamefont{Heussinger and Frey}(2006)}]{heussinger2006floppy}
\bibinfo{author}{\bibfnamefont{C.}~\bibnamefont{Heussinger}} \bibnamefont{and}
  \bibinfo{author}{\bibfnamefont{E.}~\bibnamefont{Frey}},
  \bibinfo{journal}{Physical Review Letters} \textbf{\bibinfo{volume}{97}},
  \bibinfo{pages}{105501} (\bibinfo{year}{2006}), \bibinfo{author}{\bibfnamefont{C.}~\bibnamefont{Heussinger}},
  \bibinfo{author}{\bibfnamefont{B.}~\bibnamefont{Schaefer}}, \bibnamefont{and}
  \bibinfo{author}{\bibfnamefont{E.}~\bibnamefont{Frey}},
  \bibinfo{journal}{Physical Review E} \textbf{\bibinfo{volume}{76}},
  \bibinfo{pages}{31906} (\bibinfo{year}{2007}).


\bibitem[{\citenamefont{Thorpe}(1983)}]{thorpe1983continuous}
\bibinfo{author}{\bibfnamefont{M.}~\bibnamefont{Thorpe}}, \bibinfo{journal}{J.
  Non-Cryst. Solids} \textbf{\bibinfo{volume}{57}}, \bibinfo{pages}{355}
  (\bibinfo{year}{1983}), \bibinfo{author}{\bibfnamefont{D.}~\bibnamefont{Head}},
  \bibinfo{author}{\bibfnamefont{F.}~\bibnamefont{MacKintosh}},
  \bibnamefont{and} \bibinfo{author}{\bibfnamefont{A.}~\bibnamefont{Levine}},
  \bibinfo{journal}{Physical Review E} \textbf{\bibinfo{volume}{68}},
  \bibinfo{pages}{25101} (\bibinfo{year}{2003}{\natexlab{c}}).


\bibitem[{\citenamefont{Landau and Lifshitz}(1995)}]{landau1995theory}
\bibinfo{author}{\bibfnamefont{L.}~\bibnamefont{Landau}} \bibnamefont{and}
  \bibinfo{author}{\bibfnamefont{E.}~\bibnamefont{Lifshitz}},
  \emph{\bibinfo{title}{{Theory of Elasticity, Course of Theoretical Physics
  vol 7}}} (\bibinfo{year}{1995}).

\bibitem[{\citenamefont{Lees and Edwards}(1972)}]{lees1972computer}
\bibinfo{author}{\bibfnamefont{A.}~\bibnamefont{Lees}} \bibnamefont{and}
  \bibinfo{author}{\bibfnamefont{S.}~\bibnamefont{Edwards}},
  \bibinfo{journal}{Journal of Physics C: Solid State Physics}
  \textbf{\bibinfo{volume}{5}}, \bibinfo{pages}{1921} (\bibinfo{year}{1972}).

\bibitem[{\citenamefont{Das et~al.}(2007)\citenamefont{Das, MacKintosh, and
  Levine}}]{das2007effective}
\bibinfo{author}{\bibfnamefont{M.}~\bibnamefont{Das}},
  \bibinfo{author}{\bibfnamefont{F.}~\bibnamefont{MacKintosh}},
  \bibnamefont{and} \bibinfo{author}{\bibfnamefont{A.}~\bibnamefont{Levine}},
  \bibinfo{journal}{Physical review letters} \textbf{\bibinfo{volume}{99}},
  \bibinfo{pages}{38101} (\bibinfo{year}{2007}).

\bibitem[{\citenamefont{Mohazzabi and Behroozi}(1998)}]{mohazzabi1998elastic}
\bibinfo{author}{\bibfnamefont{P.}~\bibnamefont{Mohazzabi}} \bibnamefont{and}
  \bibinfo{author}{\bibfnamefont{F.}~\bibnamefont{Behroozi}},
  \bibinfo{journal}{Langmuir} \textbf{\bibinfo{volume}{14}},
  \bibinfo{pages}{904} (\bibinfo{year}{1998}).

\bibitem[{\citenamefont{Chaikin and Lubensky}(2000)}]{chaikin2000principles}
\bibinfo{author}{\bibfnamefont{P.}~\bibnamefont{Chaikin}} \bibnamefont{and}
  \bibinfo{author}{\bibfnamefont{T.}~\bibnamefont{Lubensky}},
  \emph{\bibinfo{title}{{Principles of Condensed Matter Physics}}}
  (\bibinfo{publisher}{Cambridge Univ Pr}, \bibinfo{year}{2000}).

\bibitem[{\citenamefont{DiDonna and Lubensky}(2005)}]{didonna2005nonaffine}
\bibinfo{author}{\bibfnamefont{B.}~\bibnamefont{DiDonna}} \bibnamefont{and}
  \bibinfo{author}{\bibfnamefont{T.}~\bibnamefont{Lubensky}},
  \bibinfo{journal}{Physical Review E} \textbf{\bibinfo{volume}{72}},
  \bibinfo{pages}{66619} (\bibinfo{year}{2005}).

\bibitem[{\citenamefont{Wyart~et~al}(2008)}]{wyart2008}
\bibinfo{author}{\bibfnamefont{M.}~\bibnamefont{Wyart}},
\bibinfo{author}{\bibfnamefont{H.}~\bibnamefont{Liang}},
\bibinfo{author}{\bibfnamefont{A.}~\bibnamefont{Kabla}} \bibnamefont{and}
  \bibinfo{author}{\bibfnamefont{L.}~\bibnamefont{Mahadevan}},
  \bibinfo{journal}{Physical Review Letters} \textbf{\bibinfo{volume}{101}},
  \bibinfo{pages}{215501} (\bibinfo{year}{2008}).

\end{thebibliography}

\end{document}